\documentclass[10pt]{article}
\usepackage{amssymb,amsmath,amsthm,amsfonts,amscd}
 \usepackage{graphicx}
\newcommand\be{\begin{equation}}
\newcommand\ee{\end{equation}}
\newcommand\bea{\begin{eqnarray}}
\newcommand\eea{\end{eqnarray}}

\newcommand{\fatalpha}{{\bf \alpha \kern -0.44em \alpha}}
\newcommand{\fatsigma}{{\bf \sigma \kern -0.54em \sigma}}
\newcommand{\tpchi}{{\bf D \kern -0.35em D}}
\newcommand{\llambda}{{\bf \lambda \kern -0.45em \lambda}}

\bibliography{plain}
\begin{document}
\title{\bf Strain effects on optical properties of linearly polarized resonant modes in the presence of monolayer graphene
}
\author{A. Alidoust Ghatar
\thanks{E-mail:
Dariush110@gmail.com}$$, D. Jahani$$
 \\
 \\$${\small Materials and Energy Research Center,
Tehran, Iran.}}


\maketitle 

\begin{abstract}
Recently, huge attention has been drawn to improve optical sensing devices based on photonic resonators in the presence of graphene. In this paper, based on the transfer matrix approach and  TE polarization for the incident electromagnetic waves, we numerically evaluate the transmission and reflection spectra for one-dimensional photonic resonators and surface plasmon resonances with strained graphene, respectively. We proved that a relatively small strain field in graphene can modulate linearly polarized resonant modes within the photonic bandgap of the defective crystal. Moreover, we study the strain effects on the surface plasmon resonances created by the evanescent wave technique at the interference between a monolayer graphene and prism.
\end{abstract}

\maketitle

\section{ Introduction}
In recent years, photonic crystals (PCs) which could act also as optical resonators have attracted considerable attention due to their interesting optical properties in controlling the propagation of electromagnetic waves [1,2]. PCs are periodic dielectric structures with different refractive indices built artificially in the form of one-, two- and three dimensions [3,4,5]. These artificial layered structures exhibit photonic spectra revealing allowed and forbidden frequency regions [6,7]. The emerged frequency bandgap, however, is the most important feature of PCs which is considered in fabricating photonic devices [8]. Interestingly, by creating a defect in the crystal one can create a frequency mode within the photonic bandgap of the structure, that is, if light with this frequency is incident upon the crystals, it is transmitted through the crystals as a localized mode with low absorption [9,10,11]. Therefore, creating and analyzing these kinds of frequency modes is of most interest in photonic applications. Moreover, defect modes could be controlled by the external parameters associated with materials covering the defect layer [12]. As a result, recently, monolayer graphene has been demonstrated to be an ideal material for covering the defect layer of the defective photonic structures [13,14,15].
\\Graphene is a 2D crystalline allotrope of carbon with a honeycomb lattice which has attracted growing interest since it was experimentally discovered for the first time in 2004 [16,17,18,19,20]. Each carbon atom in graphene has four bonds, one $\sigma$ bond with each of its three neighbors, and one $\pi$-bond that is oriented out of its plane which is formed by carbon atoms with $1.42$ ${\AA}$ apart in a hexagonal lattice [21]. This hexagonal lattice which can be regarded as two interleaving triangular lattices could be stretched to some extent in order to enhance its electronic and optical properties [22,23]. This perspective has been successfully used for evaluating the electronic band structure of single-layer graphene using a tight-binding approximation [24]. Therefore, the honeycomb structure of graphene can be deformed by strain to enhance its optical properties [25,26].

Strain in graphene's lattice can be then utilized to generate various basic elements for all-graphene electronic and nano-optic applications [27,28]. Note that the growth of graphene on substrates with a different lattice constant usually introduces strain which can be detected by Raman spectroscopy [29]. Strain has also showed to affect the various electronic and optical properties that may be described by linear response correlation functions and it still arises along the edges manifesting some interesting quantum features in graphene such opening of a Dirac gap at the Fermi level. Interestingly, such effects critically depend on the direction of the applied strain [30,31,32]. In this paper, we consider the effect of uniaxial strain on the optical properties of a 1D PC with a defect covered by two graphene sheets.

Here, we should emphasize that PCs are photonic resonators for which there are different methods to be fabricated like evanescent lithography [33, 34]. However, in the absence of a PC to amplifies the signal, one can use more natural resonances of surface plasmons (SPs) which are defined as the resonant oscillation of electrons at the interface stimulated by incident light and propagating parallel to the interface. To generate SPRs with the evanescent wave (EW) the
refractive index of the first medium must be greater than the second one's. Therefore, one can directly place a graphene sheet on a prism in order to investigate the emergent SPRs. Upon resonance condition, a dip at a particular wavelength and angle of incidence would appear in the reflection spectrum. SPRs could be employed for numerous sensing applications [35, 36]. Now, the strain effect on the surface plasmonic waves could be investigated in this regard which are of most interest in sensing applications.

The paper is organized as follows. We introduce the theory and formalism in Sec. 2. Then. in Sec. 3, we present the numerical results concerning the transmission spectra of the proposed device with two graphene sheets subjected to the relativity small uniaxial strain. The effect of strain in the plasmonic wave is expressed in Sec. 4. Finally, the conclusion is addressed in Sec. 5.
\section{Formalism and Simulations}
In optics for computing the electromagnetic propagation in an infinity extended of a periodical slap and the reflection and transmission spectra, there is a useful and powerful mathematical method known as the Transfer matrix method.
Below is described how the transfer matrix is applied to electromagnetic waves (for example light) of a given frequency propagating through a stack of layers at normal incidence [37].
\\At first we consider the transmission and reflection of $EM$ waves incident on a single dielectric layer. In general, the solution of Maxwell's equations in layer $l$ can be expressed accordingly:
\begin{eqnarray}
 E(x|\omega)=E_{l1}e^{ik_{l}x}+E_{l2}e^{-ik_{l}x}.
  \end{eqnarray}
 Here $k_{l}=\frac{\omega}{c}\sqrt\varepsilon_{l}$ is the wave vector in the $lth$ layer and $c$ and $\varepsilon$ are the speed of light in vacuum and dielectric constant, respectively. The field in the medium is divided into two components, the transmitted component $E_{l1}$  and the reflected component $E_{l2}$. The coefficients $E_{l1}$ and $E_{l2}$ have to be determined from the boundary conditions that both the electric field and its first derivative are continuous across an interface. For convenience, the field is written in the form of a vector as below:
\begin{eqnarray}
 E(x|\omega)=
 \begin{pmatrix}
  E_{l1}e^{ik_{l}x} \\
 E_{l2}e^{-ik_{l}x}
\end{pmatrix}
  \end{eqnarray}
  It can be shown the field at $x_{l}$ in layer $l$ is related to the field at $x_{l-1}$ in layer $l-1$ by a $2\times2$ transfer matrix $T(x_{l-1},x_{l})$
\begin{eqnarray}
 E(x_{l}|\omega)=T(x_{l-1},x_{l})E(x_{l-1}|\omega).
  \end{eqnarray}
  where the transfer matrix is given by
\begin{eqnarray}
 T(x_{l-1},x_{l})=P_{l}(\Delta x_{l}) Q_{l-1,l} P_{l-1}(\Delta x_{l-1}).
  \end{eqnarray}
Here $\Delta x_{l}=x_{l}-d_{l-1,l}$ and  $\Delta x_{l-1}= d_{l-1,l}-x_{l-1}$ are the distance from $x_{l}$  and $x_{l-1}$ to the interface between layers $l-1$ and $l$ located at $x_{l}=d_{l-1,l}$ , respectively. It is easy to show that the matrices $P$ and $Q$ are given by
\begin{equation}
 P_{l}(\Delta x)=
 \begin{pmatrix}
  e^{ik_{l}\Delta x} & 0 \\
  0 & e^{-ik_{l}\Delta x}
\end{pmatrix},
  \end{equation}
 where, in the absence of the graphene sheets yields:
 \begin{equation}
 Q_{l-1,l}=\frac {1}{2}
 \begin{pmatrix}
  1+\eta_{l-1,l} & 1-\eta_{l-1,l} \\
  1-\eta_{l-1,l} & 1+\eta_{l-1,l}
\end{pmatrix},
  \end{equation}
while for the defect layer surrounded by graphene regarding the EM polarization for which the magnetic field is along with the y directions, it could be expressed as:
  \begin{equation}
 Q_{l-1,l}=\frac {1}{2}
 \begin{pmatrix}
  1+\eta_{l-1,l}+\xi_{l-1,l} & 1-\eta_{l-1,l}-\xi_{l-1,l} \\
  1-\eta_{l-1,l}+\xi_{l-1,l} & 1+\eta_{l-1,l}-\xi_{l-1,l}
\end{pmatrix},
  \end{equation}
 where $\eta_{l-1,l}=k_{l-1}/k_{l}$ and $\xi_{l-1,l}=\sigma k_{l}/\varepsilon_{0}\omega$ that $\sigma$ and $\varepsilon_{0}$ are the vacuum permittivity and the optical conductivity of graphene sheet in presence of uniaxial strain [38], respectively.\\

\begin{eqnarray}
 \sigma_{g}(\omega)=\frac{e^{2}}{4\hbar}\Big\{\frac{i}{2\pi}\frac{16k_{B}T}{\hbar\omega}\ln\Big(2\cosh\big(\frac{\mu}{2k_{B}T}\big)\Big)
+\frac{1}{2}+\frac{1}{\pi}\arctan\frac{\hbar\omega-2\mu}{2k_{B}T}-\frac{i}{2\pi}\ln\frac{(\hbar\omega+2\mu)^{2}}
{(\hbar\omega-2\mu)^{2}+(2k_{B}T)^{2}}\Big\},
  \end{eqnarray}
where $e$, $k_{B}$, $\mu$, and $\hbar=\frac{h}{2\pi}$ are electron charge, Boltzmann constant chemical potential and the reduced Planks constant, respectively.
Therefore, optical conductivity of strained graphene under linear regime is computed by:
\begin{eqnarray}
 \sigma_{strain}(\omega)=\sigma_{g}(\omega)[1-2\beta(1+v)s\cos(2\Theta-2\phi)].
  \end{eqnarray}

Here $\sigma_{g}$ , $\phi$ , and $\Theta$ is the optical conductivity of graphene, the orientation of the electric field and the strain angle, respectively. Where $s$, $v$ and $\beta$ are referred to as the intensity of strain, hydrostatic limit and the nearest-neighbor hopping parameter relative to unstrain graphene, respectively and are equal $v=0.14$ and $\beta=1.1$ [39, 40].

The physical meaning of the two matrices is that P propagates the electric field a distance $\Delta x$ in a uniform medium, whereas Q makes the electric field from one side of an interface to the other.\\
It is easy to show that the transfer matrix for a $(AB)^{N}$ structure, can be obtained by the following equation:
\begin{eqnarray}
 T=Q_{1\rightarrow2} P(d_{1,2}) Q_{2\rightarrow3} P(d_{2,3})\cdot\cdot\cdot P(d_{N-1,N})Q_{N\rightarrow N+1}.
  \end{eqnarray}
Where $T$ is the element of the $2\times2$ transfer matrix and $N$ is the number of layers.
Finally, the transmittance $(T)$ and reflectance $(R)$ can be calculated by using the matrix elements $T_{11}$  and $T_{12}$ as follows [41,42]:
\begin{equation}
R=\Big|\frac{T_{21}}{T_{11}}\Big|^2   \hspace{10mm}   T=\Big|\frac{1}{T_{11}}\Big|^2.                                                                                                                                 \end{equation}
\section{Numerical Results}
In this section, we consider the transmission spectra of the proposed PC with a defect layer coated by graphene layers from both sides. The structure of the device is schematically represented in Fig. 1.  In our calculation, the frequency region is ranging $4$ to $7 \ THz$. The structure of the PC is as $[air/(Si/SiO_{2})^{N} SiC(SiO_{2}/Si)^{N}/air]$, for which $A$ and $B$ represent the dielectric materials, and D is the defect layer, respectively. We choose $Si$ for the $A$, $SiO_{2}$ for the material $B$ and $SiC$ for the defect layer with $N$ the period of crystal to be $N=10$. The geometrical parameters of the photonic structure are such that the thickness of $A$ and $B$ layers are chosen as $d_{A}=d_{B}=5$ $\mu m $ and the thickness of the defect layer is taken $D=1.5\times(A+B)$. Note that, we neglect the thermal expansion of $Si$, $SiO_{2}$ and $SiC$ layers in our computation. The dielectric constant of layers for $Si$, $SiO_{2}$ and $SiC$ are $10.9$, $5.06$ and $4.4$, respectively.

At first, we compute the transmission spectra of the proposed structure without strain. It is seen from Fig. 2 that a defect mode centered at $5.479\ THz$ (red dotted line) is appeared within the corresponding photonic bandgap approximately ranging from $4.71$ to $6.046 \ THz$ at $T=10\ K$. By increasing the temperature to $T=300\ K$ the frequency position of the mode is fixed. However, in this case, the amount of transmission would be diminished (violet solid line). Interestingly, this defect mode could be shifted to a lower frequency region by applying the strain on graphene. As one can see, for $\mu=0.2\ eV$ and $s=0.2$ (considering the zigzag chain i.e $\Theta=0$) the position of defect modes would shift to the lower frequency region (blue dashed line).

In the following, we assume that the applied strain is along the zigzag chain. To analyze the situation in more detail, we show the frequency position variation of the defect mode as a function of the orientation of the electric field, $\phi$, ranging from $0$ to $\pi/2$ in Fig. 3. It should be mentioned that by increasing $\phi$ the position of the defect mode would shift to the higher frequency regions. Moreover, stronger stretching causes the position of the defect mode to move toward the lower frequency positions.

To proceed, in Fig. 4, the modulation of defect mode within the frequency bandgap for three values of strain $s=0.1$, $0.15$ \ and \ $0.2$ are considered in the constant chemical potential, $\mu=0.2/ eV$. As it is seen, stronger stretching shifts the defect mode toward the lower frequency region. Moreover, the slope of shift's variations as a function of the deviation angle $\phi$  ranging from $\phi=25$ to $\phi=65$ is considerably high. Interestingly, it is observed that in general for all values of strain there is no shift in position for the defect modes $\phi=\pi/4$.

Note that as the optical conductivity of the graphene layer can be tuned by the chemical potential, therefore the defect mode can be controlled by an external electric field or a gate voltage. In this regard, in Fig. 5, the effect of increasing chemical potential ranging from $\mu=0.2 \ eV$ to $\mu=0.6 \ eV$ for the two different intensity of strain $s=0.1$ Fig. 5 (a) and $s=0.2$ Fig. 5 (b) are presented. It reveals that considering a high value for the chemical potential leads to the blue shifting frequency. Furthermore, the slope of shifting, as to be expected, is higher for stronger applied strain in comparison with the lower stretching.

At this point, we turn our attention to the stronger deformations in graphene's lattice along armchair edges, $\Theta=\pi/6$. We illustrate the effect of armchair stretching for two different values of strains $s=0.1$  and  $0.3 $ in Fig. 6 (a) and (b), respectively. Fig. 6 (a) represents the results for $s=0.1$ as a function of $\phi$. In this case, the modulation position of the defect mode does not reveal a smooth behavior. It is clear that for $\phi=0$ to $\phi=\pi/6$  the position of the defect mode under armchair strain moves toward lower frequencies. Then, this trend is reversed for the larger angles so that the position of the mode starts toward move toward the higher frequencies. In fact, by enhancing the chemical potential the shape of curves become deeper. In other words, the ratio of modulations is increased for the high values of chemical potential. However, Fig. 6 (b) shows that the minimum value for the position of mode occurs at a lower point compared to the lower value for strain.

Finally, we try to see the obtained result as shown in Fig. 6 from another point of the view in fig. 7 (a) and (b). Here, the strain value is fixed at $s=0.1$ and $s=0.3$ and the figures display the effect of increasing chemical potential as a function of $\phi$. It is evident that from the figure that by applying armchair strain,  $\theta=\pi/6$, for three different values of chemical potential, it is evident that they have the same value in $\phi=75$ compare to the zigzag strain, $\theta=0$, which has occurred at $\phi=\pi/4$. We also can conclude that by increasing the intensity of strain, the ratio of changes is getting rather greater and the shape of the curve become narrower and sharper.

Now, we investigate the effect of strain in graphene's lattice on the reflection spectra of evanescent waves in the absence of a 1D photonic resonator. Our proposed model which is schematically shown in Fig. 8 has the Kretschmann configuration; Prism/Graphene/Analyte [43]. In the absence of the photonic resonator, a more natural resonant mode which is the oscillation of Dirac electrons of graphene could be emerged. In this case, SPRs could be created for the strained graphene which is directly placed on a prism. Here, the refractive index of the prism to create SPRs is $n=1.5$. The reflection spectrum for the proposed model in the absence of a strain field is illustrated in Fig. 9 for two different temperatures $T=300\ K$ and $T=10\ K$. Note that, as shown in this figure, the existence of SPRs corresponds to a dip in the associated reflection spectrum. Particularly, a dip is detected at $\theta=42.06$ which is greater than the critical angle ($\theta=41.81$) in the case of $f=0.025\ THz$ and $\mu=1\ meV$ at $T=300\ K$. However, it is observed that no SPRs would emerge at the low temperature, $T=10\ K$. Interestingly, as it is clear from Fig. 10, the small stretching of graphene's lattice changes both the angular position and the dip of SPR. For two different values of strain $s=0.1$  and $s=0.3$ with $\phi=60$ and $\phi=85$, when the strain intensity and $\phi$ is increased, the angular position shift to the lower angular position and become deeper and narrower.

Furthermore, in the following, the angular position of this dip for different values of the orientation of the electric field $\phi$ for applying strains $s=0.1$\ and \ $s=0.3$ along zigzag and armchair chain which is indicted by $\Theta=0$ and $\Theta=30$, respectively,  are shown in Fig. 11. It is obvious that increasing $\phi$, however, makes the dip deeper and narrower. Note that, the shifts in the angular position of the SPRs dip show to be small.

\section{Conclusion}
In this work, we have studied the strain effects in graphene's lattice on the resonant modes of a 1D PC with a defect layer coated with two graphene layers. The numerical calculations have been carried out based on the transfer matrix method in the terahertz region of light's spectrum. It was shown that both intensity and direction of the uniaxial strain could affect the properties of resonant signal in the forbidden frequency region of the proposed photopic resonator. To be more specific, it was observed that the intensity of the strain value could shift the position of the defect mode to a lower region in the frequency space within the PBG. Also, the effect of increasing the chemical potential on the defect modes was investigated which indicated that the emergent modes move toward higher frequencies.

In the absence of a photonic resonator, we simulate the effect of stretching the honeycomb lattice on emergent SPRs at the interface between a graphene sheet and a prism (see fig. 8). The results show that strain could significantly influence the properties of the dip in the reflection spectrum of the model.

\section{Data availability}
Data sharing is not applicable to this article as no new data were created or analyzed in this study.


\begin{figure}
\begin{center}
\includegraphics[width=18cm]{figure1}
\includegraphics{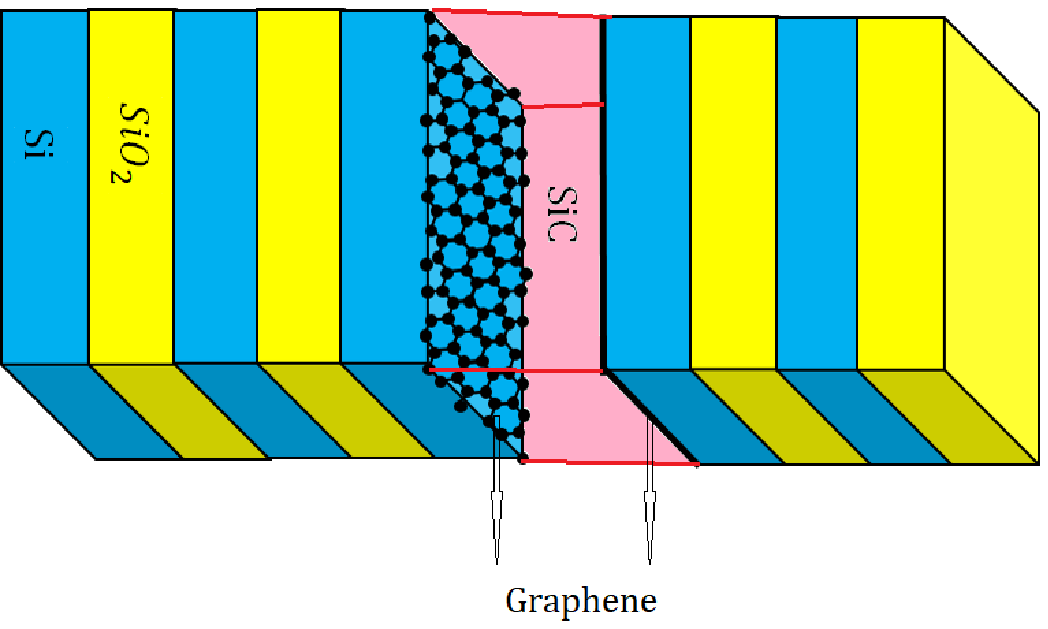}\vspace{5.3cm} \caption{The schematic representation of 1D photonic structure with a defect layer surrounded by graphene.}
\end{center}
\end{figure}

\begin{figure}
\begin{center}
\includegraphics[width=18cm]{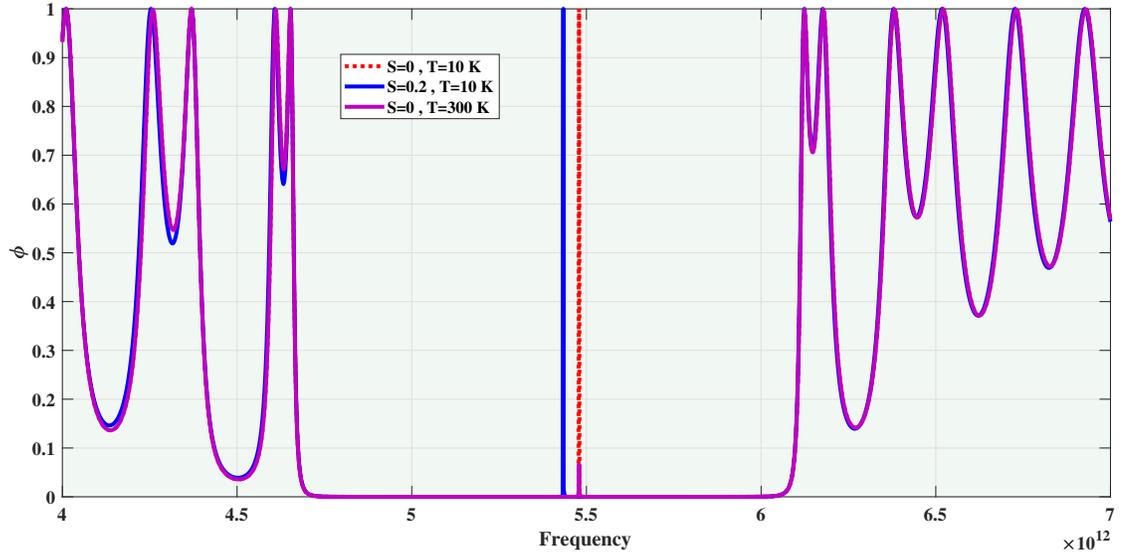}
\caption{ The frequency position of the defect mode inside the photonic bandgap of the prposed optical structure for $\mu=0.2 \ eV$ ,$\phi=0$, $\theta=0$ and $s=0$ at $T=10\ K $(red dotted line) and $T=300\ K$ (Violet solid line) and strain $s=0.2$ (blue solid line). }
\end{center}
\end{figure}

\begin{figure}
\begin{center}
\includegraphics[width=18cm]{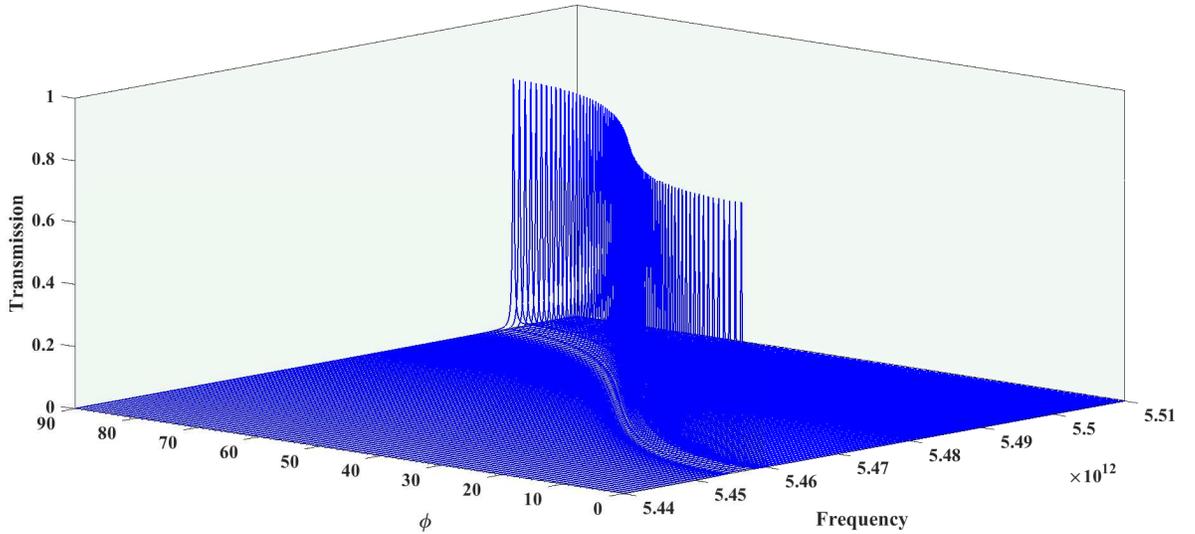}
\caption{ The transmission spectra of the defect mode as a function of orientation of the electric field $\phi$ ranging from $0$ to $\pi/2$ under strain $s=0.1$, $\mu=0.2 \ eV$ for zigzag strain, $\theta=0$. As it is also seen the frequency position of the defect mode by increasing $\phi$ move to lower frequency region.}
\end{center}
\end{figure}

\begin{figure}
\begin{center}
\includegraphics[width=18cm]{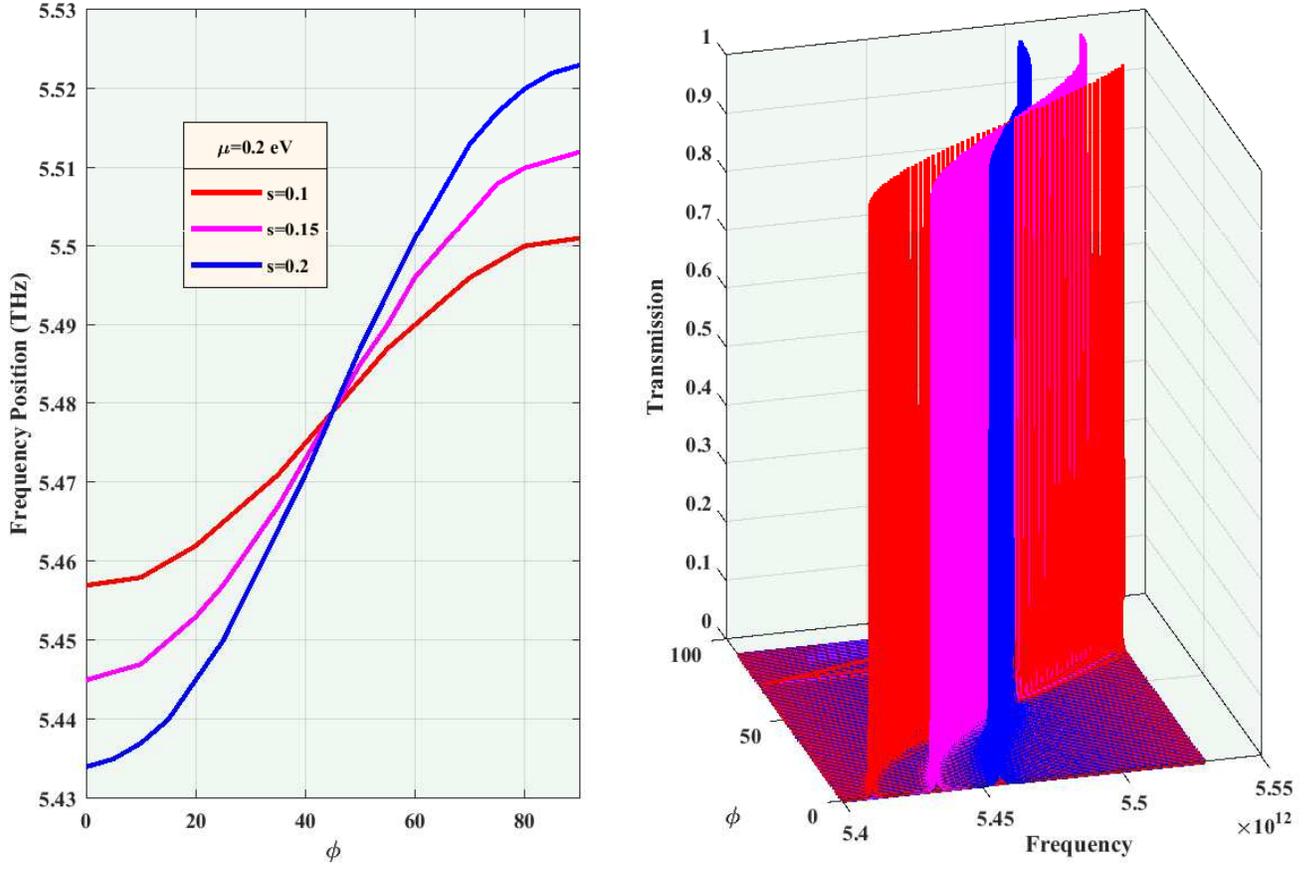}
\caption{ The schematic representaion of transmission spectra for frequency positin variation of the defect mode at $\mu=0.2 \ eV$ under three different values of strain $s=0.1$ ,  $0.15$  and  $0.2 $ for $\theta=0$.}
\end{center}
\end{figure}

\begin{figure}
\begin{center}
\includegraphics[width=18cm]{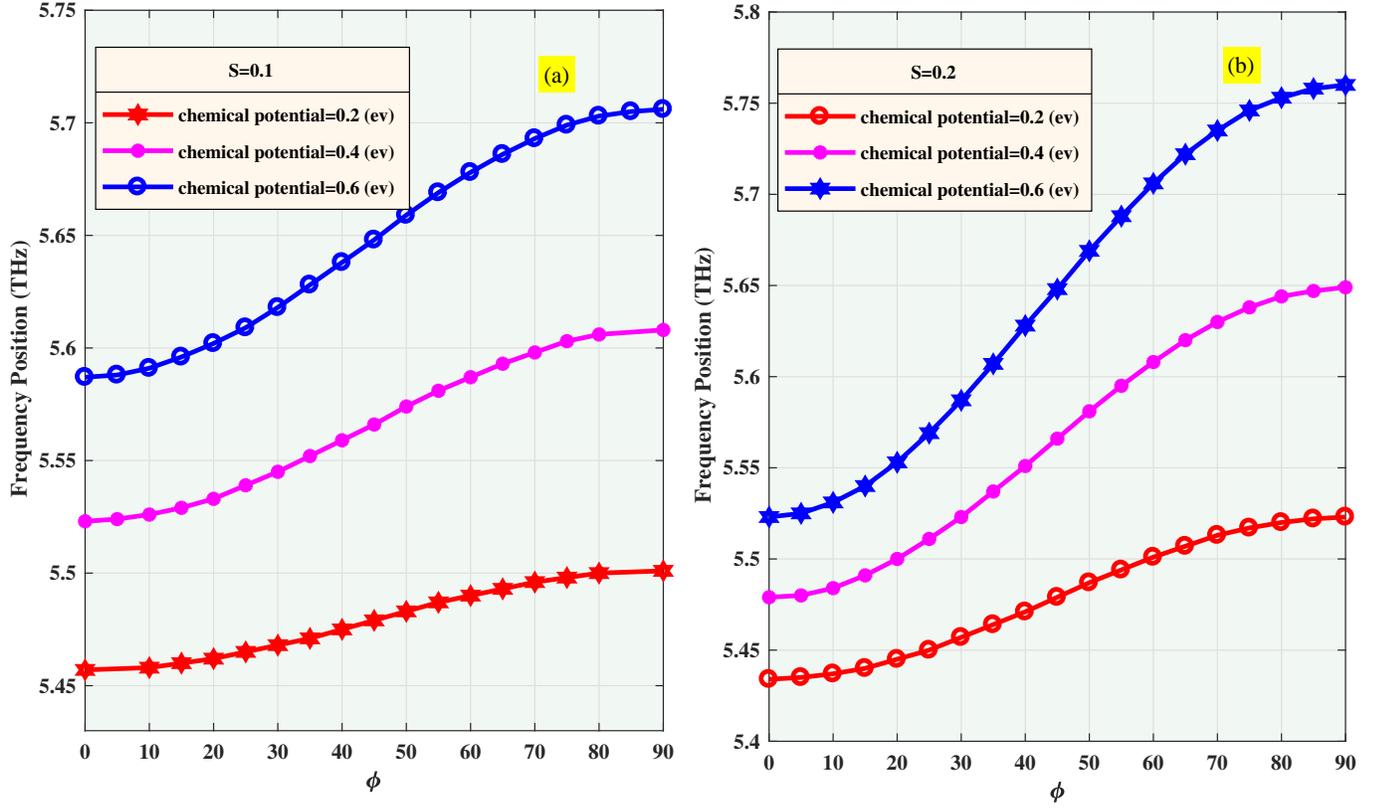}
\caption{Showing that the defect mode frequency positin can be tuned by enhancing the chemical potential $\mu=0.2$ , $0.4$ and  $0.6 \ eV$ under constant strain (a) $s=0.1$ and (b) $s=0.2$ for armchair strain, $\theta=0$, as a function of $\phi$ rengign from $0$ to $\pi/2$. }
\end{center}
\end{figure}

\begin{figure}
\begin{center}
\includegraphics[width=18cm]{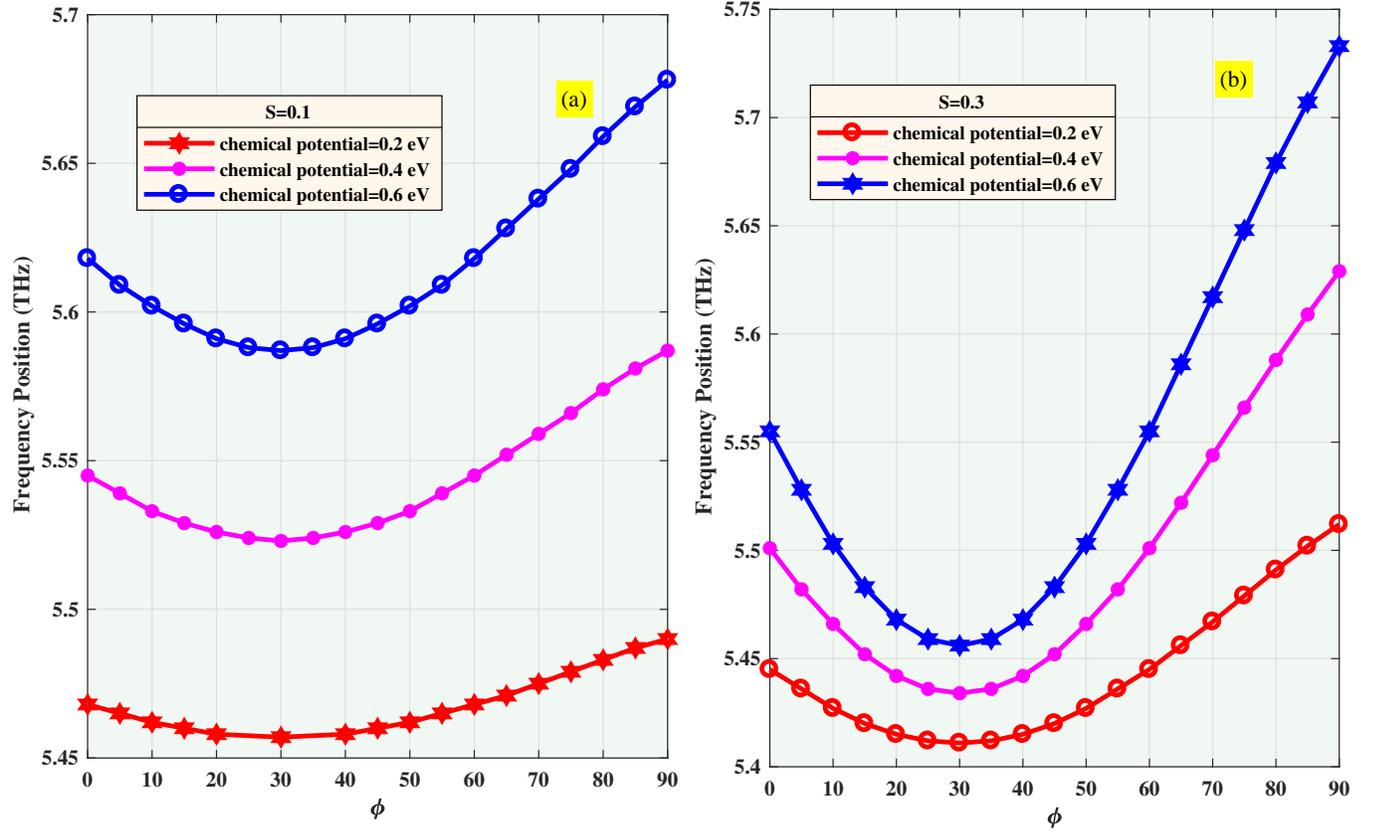}
 \caption{ The effect of increasing the direction of electric field from stretching orientation or armchair strain $\theta=\pi/6$ for the chemical potential ranging from $0.2 \ eV$ to $0.6 \ eV$ for (a) $s=0.1$ and (b) $s=0.3$. }
\end{center}
\end{figure}

\begin{figure}
\begin{center}
\includegraphics[width=18cm]{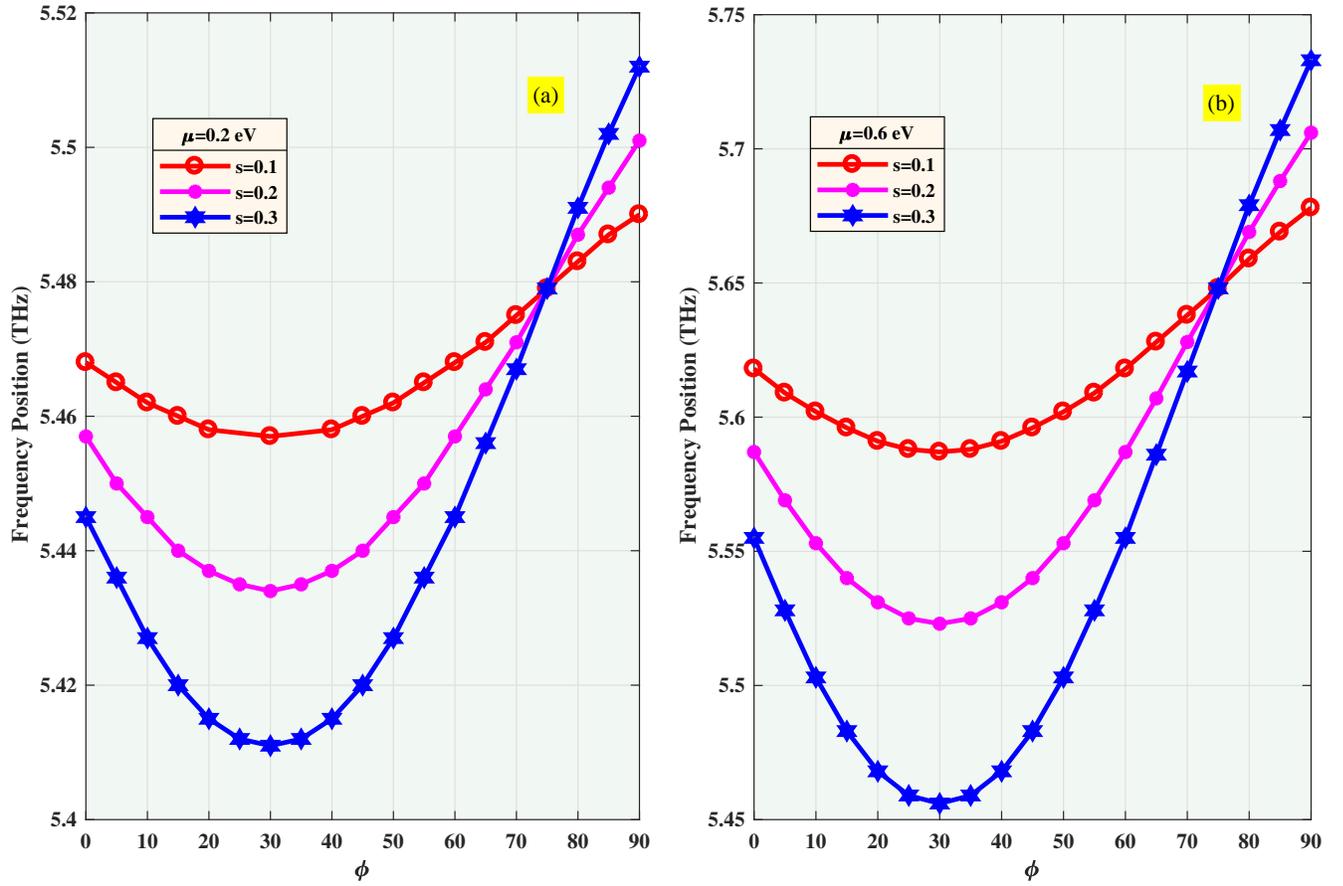}
\caption{The effect of increasing strain on the defect mode frequency position for armchair strain $\theta=\pi/6$ for (a) $\mu=0.2 \ eV$ and (b) $\mu=0.6 \ eV$}
\end{center}
\end{figure}

\begin{figure}
\begin{center}
\includegraphics[width=18cm]{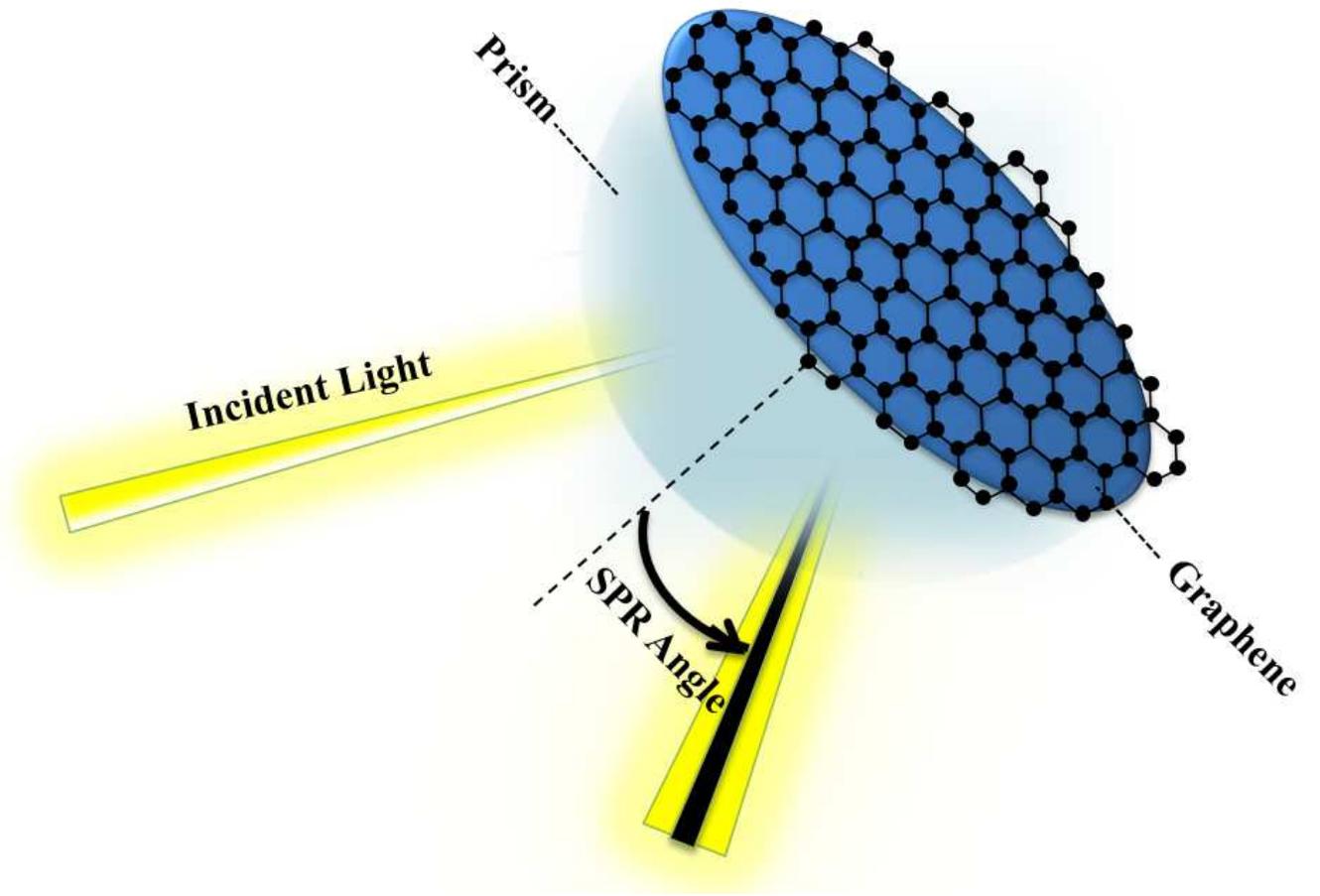}
\caption{Schematic diagram of the designed structure in the presence of a prism with $n=1.5$ coated by a single-layer graphene.}
\end{center}
\end{figure}

\begin{figure}
\begin{center}
\includegraphics[width=18cm]{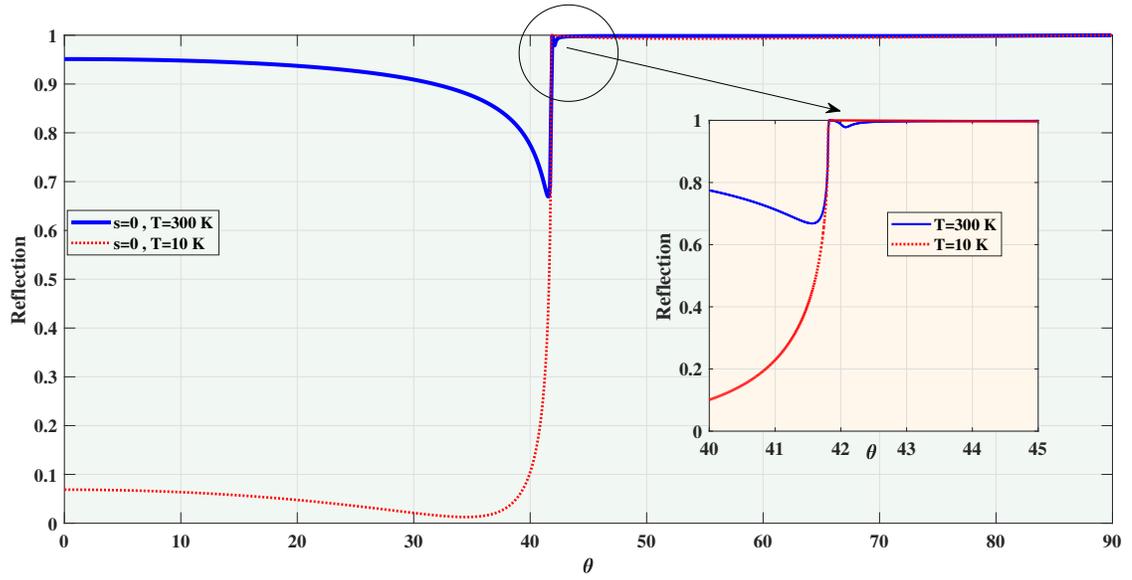}
\caption{Reflection spectrum for the SPRs of the suggested sensing structure as a function of the incident angle at $T=300\ K$ blue solid line and $T=10\ K$ red solid line.}
\end{center}
\end{figure}

\begin{figure}
\begin{center}
\includegraphics[width=18cm]{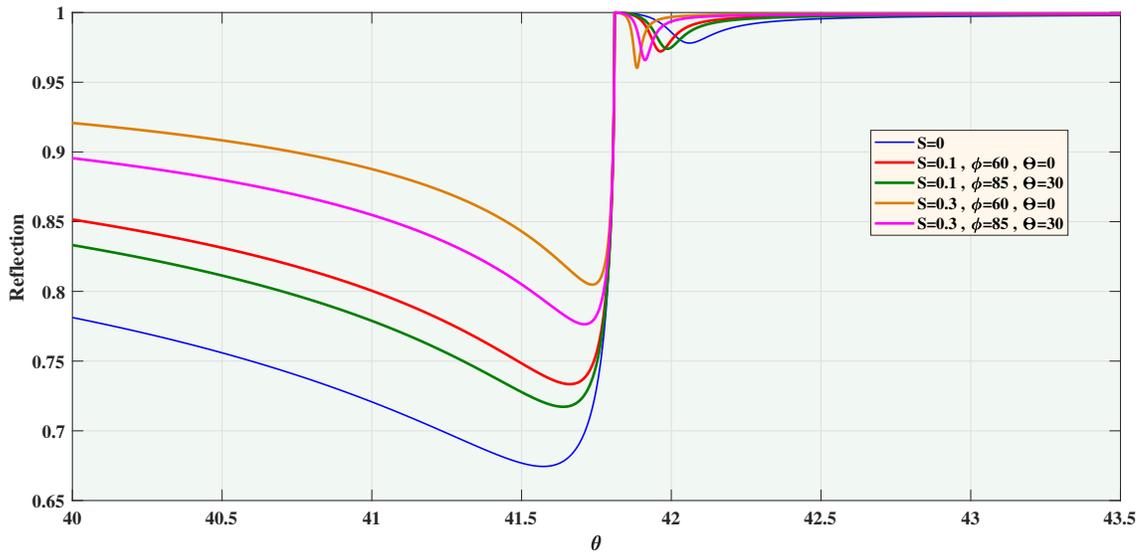}
\caption{Variations of reflectance as a function of the $\theta$ for different value of strain $s=0.1$ / and $0.2$.}
\end{center}
\end{figure}

\begin{figure}
\begin{center}
\includegraphics[width=18cm]{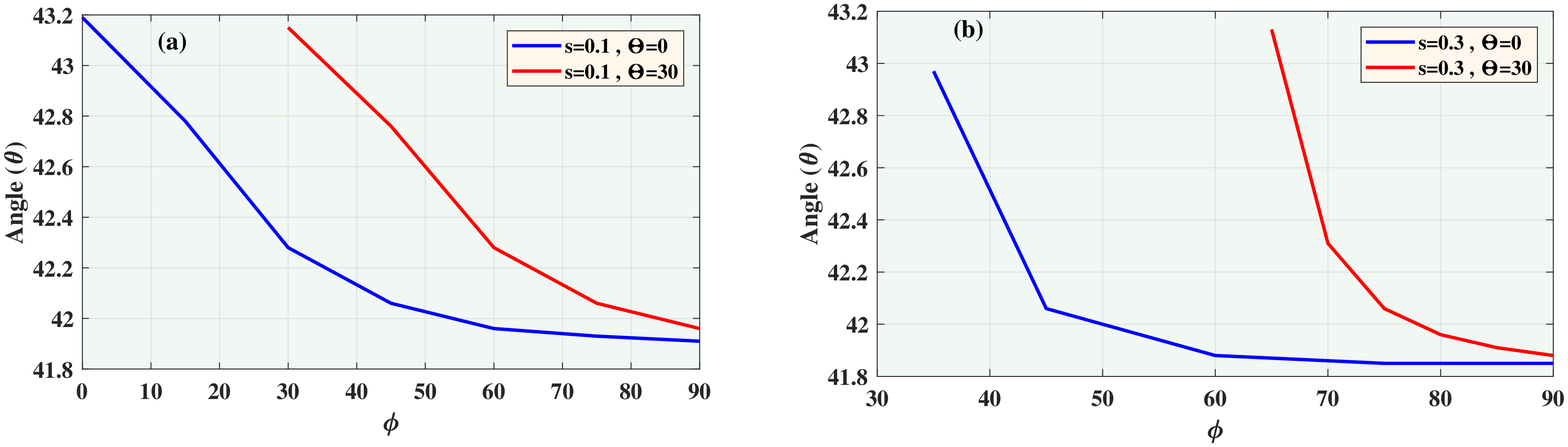}
\caption{The angle variations $\theta$ as a function of the $\phi$ under different strain (a) $s=0.1$ (b) $s=0.3$ for zigzag chain $\Theta=0$ and armchair edge $\Theta=30$}
\end{center}
\end{figure}

\end{document}